\documentclass[twocolumn,aps,pra,showpacs,amsmath,amssymb,superscriptaddress]{revtex4-1}

\usepackage{graphicx}
\usepackage{dcolumn}
\usepackage{amsmath}
\usepackage[hypertex]{hyperref}
\usepackage{amsfonts}
\usepackage{amsthm}
\usepackage{amssymb}
\usepackage{mathrsfs}
\usepackage{graphicx}
\PassOptionsToPackage{caption=false}{subfig}
\usepackage{subfig}
\usepackage{color}

\begin{document}
\title{Generation of Superposition States and Charge-Qubit Relaxation Probing in a Circuit}

\author{O. P. de S\'a Neto}
 \email{opsn@ifi.unicamp.br}
\affiliation{Instituto de F\'{\i}sica Gleb Wataghin, Universidade
Estadual de Campinas,  P.O. Box 6165, CEP 13083-970, Campinas,
S\~{a}o Paulo, Brazil}

\author{M. C. de Oliveira}
 \email{marcos@ifi.unicamp.br}
\affiliation{Instituto de F\'{\i}sica Gleb Wataghin, Universidade
Estadual de Campinas,  P.O. Box 6165, CEP 13083-970, Campinas,
S\~{a}o Paulo, Brazil}

\author{A. O. Caldeira}
\affiliation{Instituto de F\'{\i}sica Gleb Wataghin, Universidade
Estadual de Campinas,  P.O. Box 6165, CEP 13083-970, Campinas,
S\~{a}o Paulo, Brazil}

\date{\today}

\begin{abstract}
We demonstrate how a  superposition of coherent states can be
generated for a  microwave field inside a coplanar transmission
line coupled to a single superconducting charge qubit, with the
addition of a single classical magnetic pulse for chirping  of the
qubit transition frequency. We show how the qubit dephasing
induces decoherence on the field superposition state, and how it
can be probed by the qubit charge detection. The character of the charge qubit relaxation process itself is imprinted in the field state decoherence profile.
\end{abstract}

 \maketitle
\section{Introduction}
 Generation of
non-classical states of the electromagnetic field has been one of
the most pursued problems in quantum optics \cite{haroche}. This
interest was reinforced in the last years due to its  potential
applicability in quantum information \cite{ref2}.  Aside from the
experimental effort to generate single photon states, the
generation of superposition of coherent states has been for a long
time a central task, with some successful attempts. Superposition
of coherent states, commonly known as Schr\"{o}dinger cat states,
were firstly generated in a superconducting microwave cavity field
interacting with flying Rydberg atoms and its  decoherence time
was measured in \cite{raibruhaprl} also employing Rydberg atoms
sequentially interacting with the field following a previous
theoretical proposal \cite{davidovich}. More recently these states
were generated in a propagating field by photon-subtraction from a
Gaussian state \cite{Grangier}. Indeed, special superpositions,
known as odd and even coherent states have well defined parities
and can have immediate application to encode quantum bits in a
robust way \cite{Mc1}. In the last few years a new technology for
coupling superconducting qubits to coplanar waveguides has been
developed to an outstanding level \cite{jm}. Many different tests
have been developed and interesting quantum optical experiments
have been implemented in superconducting circuits \cite{rev}.
Then, a simple question could be posed on how to generate a
superposition of coherent states in this kind of system, and
subsequently how to probe its decoherence due to dissipative
effects of the qubit or the field. To our knowledge the only
proposals in that direction employ a SQUID-type two-level system
coupled to a microwave field \cite{liu}, and  a micromechanical
resonator coupled to a Cooper-Pair Box \cite{armour} (See Ref.
\cite{liu} for a complete list of connected proposals in distinct
technologies). However none of those proposals bear a resemblance
with the neat experiment in \cite{raibruhaprl}.
\begin{figure}[!ht]
\begin{center}
\setlength\fboxsep{0.5pt} \setlength\fboxrule{0.5pt}
\subfloat[]{\label{fig1}\includegraphics[scale=0.5]{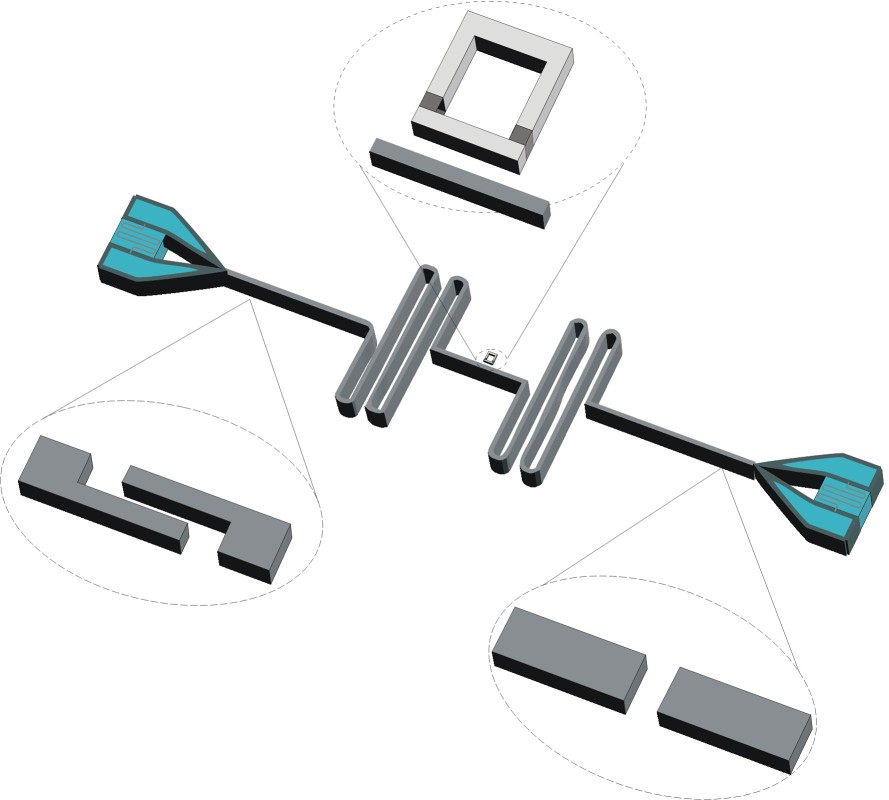}} \quad
\subfloat[] {\label{fig2}\includegraphics[scale=0.3]{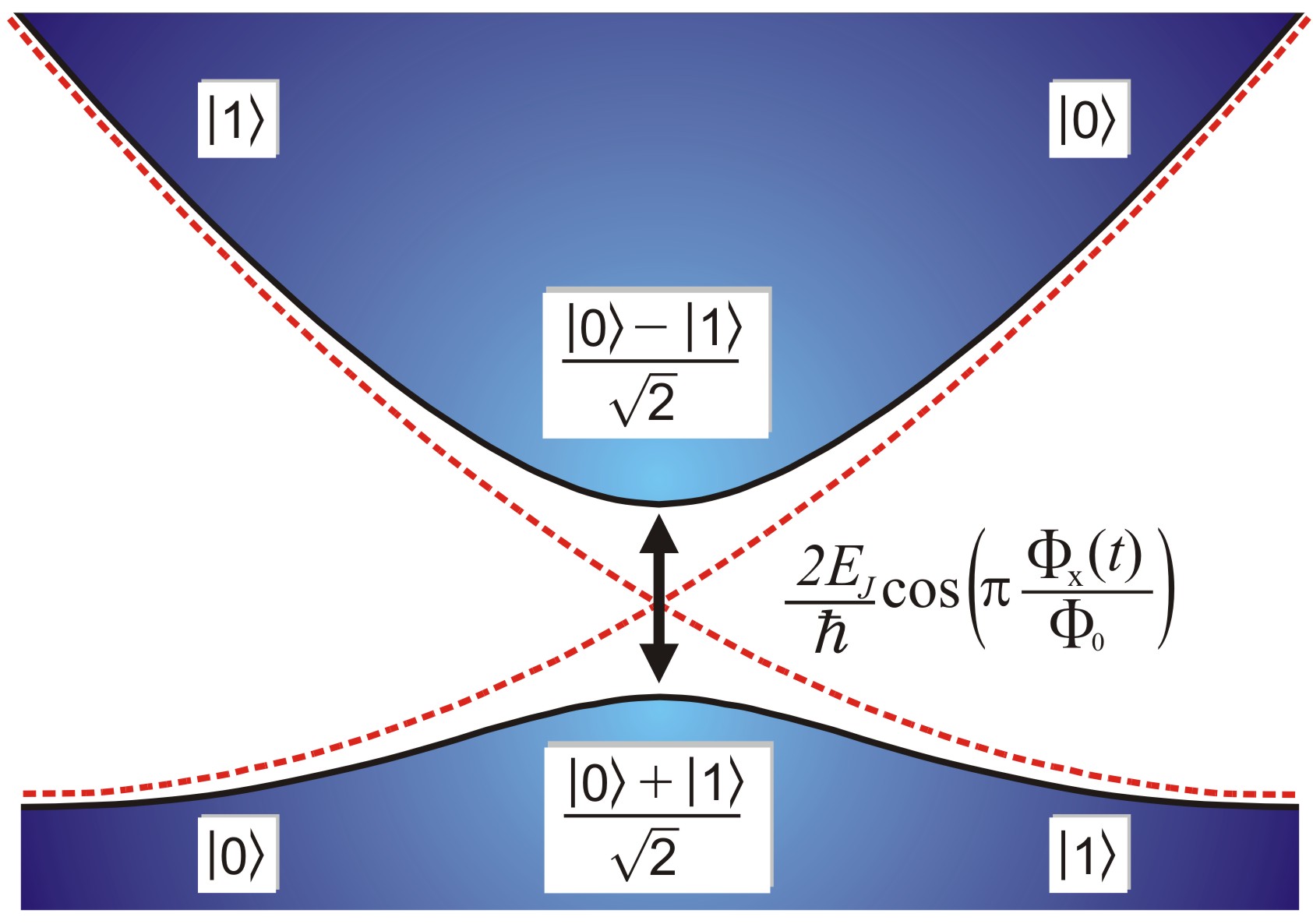}}
\end{center}
\caption{(a) Schematic setup, with the central transmission line
(resonator) capacitively coupled to the source and drain, and
capacitively coupled to a SQUID-type qubit. (b) Variable energy
levels of the qubit ($0<n_{g}<1$), with an external classical
magnetic flux $\Phi_x(t)$. The resonator field is always blue
detuned from that transition.} \label{qedc}
\end{figure}

In this paper we show how to proceed to generate a superposition
of  coherent states of a microwave field in a transmission line
resonator through the interaction with a single superconducting
charge qubit controlled by an external  single classical magnetic
pulse. In contrast to the proposal in Ref. \cite{liu}, which
employs of a  switchable interaction between the qubit and the
microwave field,  we assume a continuously varying magnetic pulse,
and show how the generation can be almost deterministic in that
way. The present proposal has the advantage to resemble the
experiment in \cite{raibruhaprl} in a realistic ground, allowing
thus for decoherence  probing through  sequential qubit
measurements and sequential classical magnetic pulses.  We show
that the qubit states dephasing is the most relevant dissipative
effect leading to decoherence of preselected field states.
Employing actual experimental parameters we show that this
proposal is compatible with present technology, and could be
actually implemented.
\section{Time dependent field-qubit coupling}
Cavity quantum electrodynamics in superconducting circuits offer a
exquisite playground for quantum information processing, and has
provided the first coherent coupling between an ``artificial
atom'', the charge qubit, and  a field mode of a resonator
\cite{mae,jm}. Mappings of qubit states \cite{1f}, and also tests
for fundamental problems, such as the Purcell effect \cite{1f} and
photon number state resolving \cite{JG} have also been achieved. The setup employed in all
those remarkable experiments is shown in Fig. \ref {fig1}, where a
niobium transmission line resonator is capacitively coupled to a
source (on the left) and to a drain (on the right). The resonator
is also capacitively coupled to the charge states of a SQUID
\cite{eeq}.   The advantage of employing a SQUID is that the
charge states can be addressed and manipulated in such a way to be
set close or far from resonance with a given resonator field mode
by an externally applied classical magnetic flux. By considering
only the ground and the first excited states near the charge
degeneracy point, the superconducting device can be well
approximated by a two level system (Fig. 1b), here addressed as a
qubit. In this regime the Hamiltonian \cite{mae} describing a quantized
electromagnetic field mode coupled to the charge qubit is given by
\begin{eqnarray}
H&=&\hbar\omega a^{\dag}a+H_{S}
\label{0}
\end{eqnarray}
{where}
 \begin{eqnarray}H_{S}&=&-\frac{1}{2}B_{z}\sigma_{z}-\frac{1}{2}B_{x}\sigma_{x},\label{s}
\end{eqnarray}{is the Hamiltonian for the qubit-field joint system, where $B_{z}\equiv4E_{C}(1-2n_{g})$,
{being $E_{C}\equiv e^{2}/2(C_{g}+C_{j})=e^{2}/2C_{\Sigma}$, the single electron charging energy with $C_{g}$ as the gate capacitance(
 associated to the accumulated charge in the island capacitively coupled
 to the central transmission line resonator) and  and $C_{j}$ the Josephson capacitance, respectively.  $n_{g}\equiv C_{g}V_{g}(t)/2e$ and  at the center of the transmission line resonator the voltage is given by $V_{g}(t)=\sqrt{\frac{\hbar\omega}{Lc}}(a^{\dag}+a)$, where $L$ is the length and $c$ is the capacitance density of the transmission line.
Here, $a$ ($a^\dagger$) is the usual annihilation (creation)
operator for the resonator field {second} mode of frequency $\omega$,
$\sigma_{x}=\left|0\right\rangle\left\langle
1\right|+\left|1\right\rangle\left\langle 0\right|$ and
$\sigma_{x}=\left|0\right\rangle\left\langle
0\right|-\left|1\right\rangle\left\langle 1\right|$, involving the
ground, $|0\rangle$, and first excited, $|1\rangle$, charge states
of the superconducting device.
$B_{x}\equiv 2E_{J}\cos\left[\pi\
\frac{\Phi_{x}(t)}{\Phi_{0}}\right]$,}  is the  Cooper-pair
tunneling energy {for two Josephson junctions}, $\Phi_{0}=hc/2e$ is the quantum of magnetic flux,
$\Phi_{x}(t)$ is a time dependent classical magnetic flux externally
applied to the SQUID. All other quantities are typical
 constants characteristic of the device \cite{eeq}. Collecting all terms the Hamiltonian, Eq. (\ref{0}) reads
\begin{eqnarray}
H=\hbar \omega a^{\dag}a+E_{J}\cos\left[\pi\
\frac{\Phi_{x}(t)}{\Phi_{0}}\right]\sigma_{x}+\hbar
g\sigma_{z}\left(a^{\dag}+a\right), \label{1}
\end{eqnarray}
being $g=({eC_{g}}/{C_{\Sigma})\sqrt{{\hbar
\omega}/{Lc}}}$  the coupling between the ``artificial atom'' (qubit) and the resonator mode. In Eq. (\ref{1}) we neglected, as usual, the term $-({e^2}/{C_{\Sigma}}) \sigma_z$ corresponding to a DC voltage shift.

 When the atomic Hamiltonian is diagonalized, the first two energy levels
 as a function of the gate charge $n_{g}\equiv C_{g}V_{g}/2e$ are
 described in fig. \ref{fig2}, where the vertical axis represents the energy
 and the horizontal represents the gate charge which is limited by the gate voltage.
Changing the basis through a rotation,
$\sigma_{z}\longrightarrow\sigma_{x}$  and
$\sigma_{x}\longrightarrow-\sigma_{z}$,  and going to the rotating
frame with the field frequency, $\omega$,  
through $R_{f}=\exp\left[i\omega t(\sigma_{z}+a^{\dag}a)\right]$,
gives
\begin{eqnarray*}
R_{af}HR_{af}^{\dag}\underbrace{R_{af}\left|\psi\right\rangle}_{\left|\psi^{'}\right\rangle}&=&i\hbar R_{af}\frac{\partial}{\partial t}R_{af}^{\dag}\underbrace{R_{af}\left|\psi\right\rangle}_{\left|\psi^{'}\right\rangle}\nonumber\\
&=&i\hbar R_{af}(\frac{\partial}{\partial t}R_{af}^{\dag})\left|\psi^{'}\right\rangle\nonumber\\
&&+i\hbar R_{af}R_{af}^{\dag}(\frac{\partial}{\partial t}\left|\psi^{'}\right\rangle)
\label{}
\end{eqnarray*}
resulting in
\begin{eqnarray*}
\underbrace{\left\{R_{af}HR_{af}^{\dag}-i\hbar R_{af}(\frac{\partial}{\partial t}R_{af}^{\dag})\right\}}_{H^{'}}\left|\psi^{'}\right\rangle&=&i\hbar\frac{\partial}{\partial t}\left|\psi^{'}\right\rangle
\label{}
\end{eqnarray*}
with the transformed Hamiltonian given by
\begin{eqnarray}
H^{'}=\sigma_{z}\left(E_{J}\cos\left[\pi\frac{\Phi_{x}(t)}{\Phi_{0}}\right]-
\hbar \omega\right)+\hbar
g\tilde{\sigma}_{x}\left(\tilde{a}^{\dag}+\tilde{a}\right)
\label{4}
\end{eqnarray}
where $\tilde{\sigma}_{x}=\sigma_{+}\exp(2i\omega
t)+\sigma_{-}\exp(-2i\omega t)$,  $\tilde{a}=a \exp(-i\omega t)$
and $\tilde{a}^{\dag}=a^{\dag} \exp(i\omega t)$, with
$\sigma_{x}=\left|-\right\rangle\left\langle
+\right|+\left|+\right\rangle\left\langle -\right|$ and
$\sigma_{z}=\left|-\right\rangle\left\langle
-\right|-\left|+\right\rangle\left\langle +\right|$. This new
Hamiltonian (\ref{4}), is analogous to the usual one for
atom-field dipole interaction. The appropriate experimental
regime is $\hbar
g<<\left(E_{J}\cos\left[\pi\frac{\Phi_{x}(t)}{\Phi_{0}}\right]-\hbar
\omega\right)<<\Delta $,  where $\Delta$ is the superconducting
energy gap.
In the following we shall consider a specific time dependent external
classical  magnetic pulse $\Phi_x(t)$ applied to the SQUID, in order to
bring the two lower states closer to resonance with the resonator
field. As we will show, during the pulse, the field accumulates a
phase conditioned to the qubit state. For that we consider the
resonator field as being always far blue-detuned from these two
lower atomic states, and so we must keep the counter-rotating
terms \cite{JG}. With that in mind we derive the formal solution
for the evolution operator
\begin{eqnarray}
U(t,t_{0})&=&1+\sum^{\infty}_{n=1}\frac{1}{(i\hbar)^{n}}\int^{t}_{t_{0}}dt_{1}
\int^{t_{1}}_{t_{0}}dt_{2}\ldots\int^{t_{n-1}}_{t_{0}}dt_{n}\nonumber\\
&&\times\tilde{V}^{I}(t_{1})\tilde{V}^{I}(t_{2})\ldots\tilde{V}^{I}(t_{n}),
\label{10}
\end{eqnarray}
with $\tilde{V}^{I}(t)\equiv
U_{0}^{\dag}\left\{g\tilde{\sigma}_{x}\left(\tilde{a}^{\dag}+
\tilde{a}\right)\right\}U_{0}$,
  and
$$ U_{0}=\exp\left[i\sigma_{z}\int_{t_{0}}^{t}dt^{'}\left(
\frac{E_{J}}{\hbar}\cos\left[\pi{\Phi_{x}(t^{'})}/{\Phi_{0}}\right]-\omega\right)\right].$$

For $k_{B}T<<E_{J}<<E_{C}<<\Delta$, a perturbative approach up to
second order in $g$ in Eq. (\ref{10}) is sufficient to describe
the dynamics of the system. For that the temperature must be as
low as $30\textrm{mK}$, which is consistent  with this kind of
experiment. In what follows we consider the solution of (\ref{10})
through time-dependent perturbation with
$H_{0}=\sigma_{z}\left(E_{J}cos\left[\pi\frac{\Phi_{x}(t)}{\Phi_{0}}\right]-\hbar
\omega\right)$ and $V_{I}^{S}=\hbar
g\left(\tilde{\sigma}^{+}+\tilde{\sigma}^{-}\right)\left(\tilde{a}^{\dag}+\tilde{a}\right)$,
with $H_{0}>>V_{I}^{S}$, since we are interested in  the radiation
field  blue detuned from the qubit transition, which on its turn
oscillates with time. For our numerical calculations, we use the
following classical magnetic flux
\begin{equation}
\Phi_{x}(t)=\frac{A\Phi_{0}}{2}\cos\left[\sigma t+\varphi\right],
\label{P}
\end{equation}
where $A=0.7$ is a strength constant, and $\sigma=8\pi \times
10^{6}\textrm{Hz}$ is the frequency of the   classic magnetic
pulse applied to the qubit in accordance with experimental values.
By further assuming typical experimental parameters to reach
$k_{B}T<<E_{J}<<E_{C}<<\Delta$,  for $T\approx30\textrm{mK}$,  we
set $k_{B}T\approx3\mu \textrm{eV}$, $E_{J}/ \hbar \approx 15,9
\times 10^{10} \textrm{Hz}$, $E_{C}=250 \mu \textrm{eV}$ and
$\Delta\approx 458,3 \mu\textrm{eV}$. The frequency of the field
in the resonator, $\omega=90\times10^{10}\textrm{Hz}$, is also
compatible with the experiments, for which the resonator quality
factor $Q=\frac{\omega}{\delta
\omega}\longrightarrow10^{4}-10^{6}$ is achievable \cite{fab}.
\begin{figure}[h]
\begin{center}
\includegraphics[angle=0,width=2.8in]{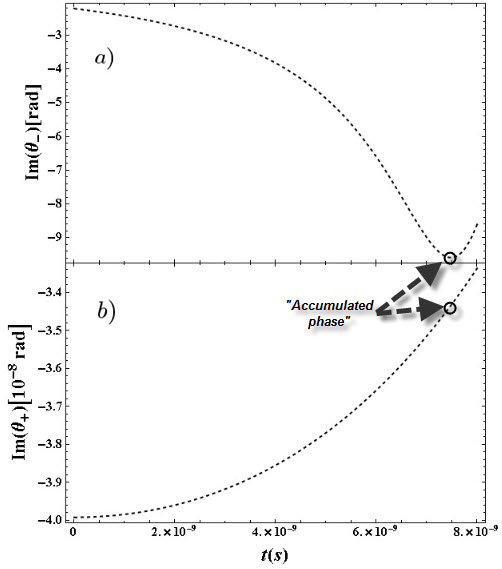}
\caption{Phase of the  resonator coherent field, due to the time
dependent  interaction with the qubit prepared in $a)$ the ground
state, and $b)$ the excited state.  The pulse $\Phi_{x}(t)$
oscillates for a half period ($\Delta t\approx 7.5 \textrm{ns}$)
with frequency $8\pi \times 10^{6}\textrm{Hz}$.} \label{fig4}
\end{center}
\end{figure}

\section{accumulated phase and conditional generation of superposition states}

In order to depict our results in a more convenient way we choose
that the field in the central line resonator is prepared in a
coherent state $|\alpha\rangle$ with an average number of photons
smaller than one, so that we can make  the following approximation
for short time,  $ 1+ \theta_{\pm}(t) a^\dagger a\approx
e^{\left[\theta_{\pm}(t) a^{\dag}a\right]}$  (see the Appendix) so
that
\begin{equation}
\left[1+ \theta_{\pm}(t) a^\dagger a \right]\left|\pm\right\rangle\left|\alpha\right\rangle\rightarrow e^{-\frac 1 2 |\alpha|^2f_{\pm}(t)}\left|\pm\right\rangle\left|\alpha
e^{\theta_{\pm}(t)}\right\rangle \label{11}
\end{equation}
where  $f_{\pm}(t)=[1-e^{2 Re (\theta_{\pm}(t))}]$, $\theta_{\pm}(t)=1+\frac{F_{\pm}(t)}{G_{\pm}(t)}$, being
\begin{eqnarray}
F_{\pm}(t)&=&\int^{t}_{0} dt_{1}e^{\mp 2i \left(\int^{t_{1}}_{0}dt^{'}\Omega(t') \right)\mp i w t_{1}}\nonumber\\
&&\times\int^{t_{1}}_{0} dt_{2}e^{\pm 2i\left(\int^{t_{2}}_{0}dt^{''}\Omega(t'')\right)
\pm i w t_{2}},\label{F}\end{eqnarray}\begin{eqnarray}
G_{\pm}(t)&=&\int^{t}_{0} dt_{1}e^{\mp 2i \left(\int^{t_{1}}_{0}dt^{'}\Omega(t') \right)\pm i w t_{1}}\nonumber\\
&&\times\int^{t_{1}}_{0} dt_{2}e^{\pm
2i\left(\int^{t_{2}}_{0}dt^{''}\Omega(t'')\right)\mp i w t_{2}},
\label{G}\end{eqnarray} with
$\Omega(t)=\omega-\frac{E_{J}}{\hbar}cos\left[\pi\frac{\Phi_{x}(t)}{\Phi_{0}}\right]$,
as  given by the second order terms from Eq. (\ref{10}).
 From Eq. (\ref{11}) it is possible to understand that as the qubit
is brought closer to resonance  with the resonator field, it will
imprint an accumulated phase on it, given by
$\textrm{Im}[\theta_\pm(t)]$ conditioned on the qubit state
$|\pm\rangle$. In Fig. \ref{fig4} we depict the numerical results
for those two conditioned accumulated phases. We see that
practically only when the qubit is in the state $|-\rangle$  the
phase in $\alpha$ is changed. With an appropriate accumulation of
$-3\pi$, as shown in  Fig. \ref{fig4}a, it is possible to create a
state $\left|-\alpha\right\rangle$ if the qubit is initially in
the $|-\rangle$ state. We have consistently checked that this
approximation is indeed very good, not only for small $\alpha$, if
we respect a balance between the field intensity and the operation
time. Moreover, we observed that around the  time of optimal phase
accumulation, $t_{op}=7.5 \textrm{ns}$,  the real part of
$\theta_\pm(t)$, related to damping or amplification, is
negligible, ($\approx 10^{-3} -10^{-4}$), so that in Eq.
(\ref{11}), $f_{\pm}(t)=0$ {at the instant of measurement as shown
in  the Appendix, and shall not be considered from now on.

To generate the superposition of
coherent states for the field, we note that due to the low
temperature of the system it is easy to prepare the qubit state
initially in $\left|0\right\rangle$. Obviously the decoupled
coherent state of the resonator and the qubit state may be written
as $
\left|0\right\rangle\otimes\left|\alpha\right\rangle={\left(\left|-\right\rangle+
\left|+\right\rangle\right)\otimes\left|\alpha\right\rangle}/{\sqrt{2}}$.
 If the aforementioned pulse is applied to the qubit, coupling it to the resonator field
 through the evolution given by  Eq. (\ref{10}), we shall have \begin{equation}
{U(t_{0}+\Delta
t,t_{0})}|0\rangle\otimes\left|\alpha\right\rangle\underbrace{\Longrightarrow}_{Pulse}
\frac{\left|-\right\rangle\otimes\left|-\alpha\right\rangle+\left|+
\right\rangle\otimes\left|\alpha\right\rangle}{\sqrt{2}}
\label{14},
\end{equation} or
$ [\left|0\right\rangle\otimes\left(\left|-\alpha\right\rangle+
\left|\alpha\right\rangle\right)+\left|1\right\rangle\otimes
\left(\left|-\alpha\right\rangle-\left|\alpha\right\rangle\right)]/{2},
\label{15} $ which is an entangled state between the qubit and the
resonator field. Consequently, the resonator field can be left in
an odd or even superposition of coherent states depending on the
detection of the qubit state with a single electron transistor
\cite{eeq}.

\section{Field state decoherence and qubit relaxation probing}

Certainly, the exact preparation of such a state is  compromised
by external noise. In contrast to experiments with microwave
fields and Rydberg atoms, dissipative effects are most noticeable
for the qubit states, which can flip from the ground state to the
excited one and vice versa due to thermal effects and inductive
coupling of the qubit to the external circuit. While the
relaxation time of the qubits are of the order of
$10^{-6}\textrm{s}$ , the relaxation of the resonator field is of
the order of $10^{-3}\textrm{s}$  and can, in principle, be
neglected. If compared with the time of the pulses for the
accumulated phase of the field, $\Delta t\approx 7.5\times
10^{-9}\textrm{s}$ , the relaxation of the atom is negligible as
well, but certainly will be important if any further manipulation
is to be executed. So in fact the effects of dissipation are  more
relevant after the pulse is applied, i.e., when $\Phi_{x}(t)=0$,
and will affect the probability to detect a given qubit state, and
consequently the generation of an appropriate field. To understand
the effects of noise in the system, we couple the qubit two-state
to a bath of harmonic oscillators in an adaptation of the standard
spin-boson model with Ohmic dissipation \cite{eeq,leg,weiss},
through the Hamiltonian
$H=H_{S}+H_{AR}+H_{R}$.
Here  $H_S$ is given by Eq. (\ref{s}),
 presented in reference \cite{eeq}, with the convention at the degeneracy point} $
\left|+\right\rangle=(\left|0\right\rangle-\left|1\right\rangle)/\sqrt{2}$,
and $
\left|-\right\rangle=(\left|0\right\rangle+\left|1\right\rangle)/\sqrt{2}$.
Eq. (\ref{s})  can be conveniently rewriten as
\begin{eqnarray}
H_{S}&=&-\frac{\Delta E}{2}(\frac{1}{\sqrt{2}}\sigma_{z}+\frac{1}{\sqrt{2}}\sigma_{x}),\end{eqnarray}
where $\Delta E=\sqrt{B_{z}^{2}+B_{x}^{2}}$, and now since $\Phi_{x}(t)=0$, $B_{z}\equiv4E_{C}(1-2n_{g})$, and $B_{x}\equiv 2E_{J}$. Thus $\Delta E =\sqrt{(2E_{J})^{2}+16E_{C}^{2}(1-2n_{g})^{2}}$, and conveniently choosing $n_{g}=0.5$ we end up with  $\Delta E =2E_{J}$. The coupling of the qubit to the bath is given by
\begin{equation}
H_{AR}=\sigma_{z}\sum_{a}\lambda_{a}x_{a},\end{equation}  and
\begin{equation}H_{R}=\sum_{a}\left(\frac{p_{a}^{2}}{2m_{a}}+\frac{m_{a}\omega_{a}^{2}x_{a}^{2}}{2}\right),\end{equation} corresponds to the bath free Hamiltonian.

The master equation for the evolution of the two level system
coupled to a bath in a thermal state
$\rho_{R}=\exp\left[H_{R}/k_{B}T\right]$ was extensively studied
in the past \cite{leg,weiss}. We  employ solutions corresponding
to an Ohmic bath \cite{eeq,leg,weiss}. By taking the previous
initial state, $|0\rangle |\alpha\rangle$, the main consequence of
the qubit relaxation will be on the probability to generate the
superposition states
$\left|0\right\rangle_{L}={(\left|\alpha\right\rangle+\left|-\alpha\right\rangle)}/{\sqrt{2}}
$, or
$\left|1\right\rangle_{L}={(\left|-\alpha\right\rangle-\left|\alpha\right\rangle)}/{\sqrt{2}}$
after the pulse is applied. To turn our description  clearer we
consider the optimal time for phase accumulation as our time
origin. Remark that since the central transmission line resonator
field is now far detuned from the qubit, they remain uncoupled.
The only possible correlation between the qubit and the field is
due to their past interaction. Taking the result for the density
operator evolution from Refs. \cite{eeq,leg,weiss} we obtain the
following joint qubit-field state
\begin{eqnarray}
\rho(t)&=&P_0(t)|0\rangle\langle 0|\otimes|0\rangle\langle 0|_L +
P_1(t)|1\rangle\langle 1|\otimes|1\rangle\langle 1|_L\label{rho}\\
&&+P_T(t)|1\rangle\langle 0|\otimes|1\rangle\langle 0|_L+
P_T^*(t)|0\rangle\langle 1|\otimes|0\rangle\langle
1|_L,\nonumber\label{r}\end{eqnarray} where $P_{0\choose 1}(t)=Tr\{|{0\choose
1}\rangle\langle {0\choose1}| \rho(t)\}$
is the probability to detect the qubit in the state $|{0\choose
1}\rangle$, given by
\begin{eqnarray}
P_{0\choose{1}}(t)&=&\frac12\left\{\tanh(\Lambda)
+[1-\tanh(\Lambda)] \exp(-t/\tau_{r})\right.\nonumber\\
&&\left.\pm\cos(\Delta E t/\hbar) \exp(-t/\tau_{\varphi})\right\},\end{eqnarray}
and  $P_T(t)=Tr\{|{0}\rangle\langle {1}| \rho(t)\}$
 is the transition amplitude, explicitly given by
\begin{eqnarray}
P_T(t)&=&-i\sin{(\Delta E t/\hbar)}
\exp(-t/\tau_\varphi),\end{eqnarray}
 where  $\Delta E=2E_{J}$, $\Lambda\equiv\Delta E/2k_{B}T$,
$\tau_{r}=[\pi\beta\Delta E \coth(\Lambda)/\hbar]^{-1}$ is the
relaxation time and $\tau_{\varphi}=[\tau_{r}^{-1}/2+2\pi\beta
k_{B}T/\hbar]^{-1}$ is the dephasing time, with $\beta\approx0.001$  a
dimensionless parameter reflecting the strength of dissipation
\cite{eeq}.  As $\Delta E\gg k_B T$,  $\tanh(\Lambda)\approx
1$, and so \begin{eqnarray}
P_{0}(t)&=&\frac12[1+\cos(\Delta E t/\hbar) \exp(-t/\tau_{\varphi})], \\
P_{1}(t)&=&\frac12[1-\cos(\Delta E t/\hbar)
\exp(-t/\tau_{\varphi})],\end{eqnarray}  reflecting the
probability to detect the qubit in the state $|0\rangle$ or
$|1\rangle$, respectively, at an instant $t$ after the classical
magnetic pulse. The entangled state from Eq. (\ref{rho}) allows a
probabilistic generation of the field superposition state, as well
as an indirect probe of the qubit state by the field state
measurement.

As can be readily checked, if the qubit detection is made right
after  the pulse is applied, there is a
high probability that the state $ \left|0\right\rangle_{L}$ will
be generated, in this case thus almost deterministically.  If the
detection is made after a long delay time,
$t\approx3\tau_\varphi$, there is 50 $\%$  chance that the qubit
can be detected in $|0\rangle$ and 50 $\%$ in $|1\rangle$ and thus
that the states $\left|0\right\rangle_{L}$ or  $
\left|1\right\rangle_{L}$, be generated, respectively.  Indeed, if
a second classical magnetic pulse is applied, after a qubit
detection, the state generated is exactly equal to (\ref{rho}) and
a sequential detection after a time interval $t$, allows the
inference of those probabilities directly from the measurement,
since the conditioned probability for two consecutive detections
in the state $|0\rangle$ is $P_{0,0}(t)=P_{0}(t)$, and the
probability for detecting $|0\rangle$ and then $|1\rangle$ is
$P_{0,1}(t)=P_{1}(t)$. \begin{figure}[h]
\begin{center}
\includegraphics[angle=0,width=3in]{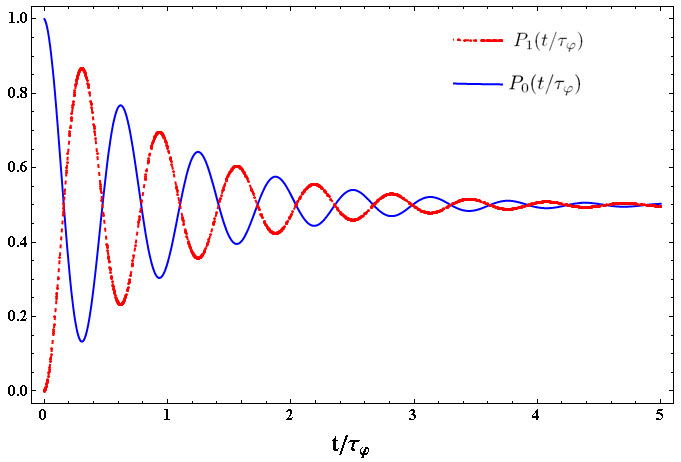}
\caption{Probabilities of charge qubit detections and consequently of postselected
$|0\rangle_L$, or $|1\rangle_L$ field states. The decoherence of the preselected
field state is given by $2P_0(t)-1$.}
\label{p001}
\end{center}
\end{figure}
This  contrasts with the decoherence time detection in Ref.
\cite{davidovich}. Those probabilities are depicted in Fig. 3, for
$T=30$ mK. There we employ an artificial $\Delta E$, which is
$10^3$ times smaller than the real one to depict the oscillatory profile.
The probability to sequentially detect $|0\rangle$  decreases with
time, as a signature of the dephasing of the  qubit states and
thus on the probability to generate  $|0\rangle_L$. Then of course
the probability to generate  $|0\rangle_L$ increases with time.
The decaying oscillatory profile depicted in Fig. 3, which
represents exactly the dephasing  of the qubit states, will surely
be imprinted in the field states even in the absence of
detections, and is a signature of the Ohmic character of the reservoir. The preselected field state derived from Eq.
(\ref{rho}) is
\begin{eqnarray}
\rho_f(t)&=&\frac1{N(t)}\left\{ |\alpha\rangle\langle \alpha|+|-\alpha\rangle\langle -\alpha|\right.\\
&&\left.+ \cos(\frac{\Delta
Et}{\hbar})e^{-t/\tau_\varphi}\left[|\alpha\rangle\langle
-\alpha|+|-\alpha\rangle\langle
\alpha|\right]\right\},\nonumber\label{rho2}\end{eqnarray} with
$N(t)=2[1+\cos(\Delta
Et/\hbar)e^{-t/\tau_\varphi}e^{-2|\alpha|^2}]$. So, the dephasing
causes the decoherence of the state $\rho_f(t)$, whose behavior is
exactly the one depicted in fig. 3, and can be inferred as being
$2P_0(t)-1$. We remark that in the same sense the decoherence of the field can be used to probe different noise characters on the qubit. Here we analyzed the effect of an Ohmic bath. Would that be a super-Ohmic or sub-Ohmic its signature would be imprinted in the field decoherence.

\section{Discussion}
We have discussed the possibility to generate superposition states
of  the field due to a classical magnetic pulse  which causes a
chirping of the frequency of the qubit, bringing it closer to
resonance with the resonator field mode. At low temperatures the
dissipative effects of the field mode are almost negligible during
the time interval the superposition is generated, while
the dephasing of the qubit becomes the
relevant source for decoherence of the state of the radiation
field. In this case the qubit detection probability affects the probability for generation of the field superposition states $|0\rangle_L$ or $|1\rangle_L$. Those states are quite
robust against dissipation \cite{Mc1}, and therefore may be quite
relevant for quantum information processing involving a hybrid
form of qubits involving field and charge qubits.  One immediate
example is the  Bell-like state Eq.(8), for
$t\ll\tau_{\varphi}$, which may directly allow the implementation
of quantum communication protocols, or indirectly through a
possible coupling to other solid state qubits \cite{dicarlo}. We leave this point
for future discussion.

\acknowledgements{We thank CAPES, FAPESP,  and CNPq for financial support
through the National Institute for Science and Technology of
Quantum Information (INCT-IQ).
}

\section*{Appendix}

\renewcommand{\theequation}{A-\arabic{equation}}
\setcounter{equation}{0}

In this appendix we show how Eqs. (\ref{11})-(\ref{G}) are
derived. For that we have to perturbatively compute how the
evolution operator,  $U(t,t_{0})$, shown in Eq. (\ref{10}) acts
over an initial state
$\left|0\right\rangle\left|\alpha\right\rangle$,
\begin{eqnarray}
U(t,t_{0})\left|0\right\rangle\left|\alpha\right\rangle=
U(t,t_{0})\frac{\left(\left|-\right\rangle+\left|+\right\rangle\right)}{\sqrt{2}}\left|\alpha\right\rangle.
\label{}
\end{eqnarray}
Due to the parity of the classical flux $\Phi_x(t)$ applied over
the qubit, Eq. (\ref{P}),  for a ``fast" pulse, i.e., a pulse with
duration equal to or shorter than half period of oscillations (for
$t_{0}=0$), the first order terms in Eq. (\ref{10}) are negligible
and we consider only terms of second order. So that,
\begin{eqnarray}
U(t,t_{0})\left|+\right\rangle\left|\alpha\right\rangle&\approx&-g^{2}\left[a^{\dag}aF_{+}(t)+aa^{\dag}G_{+}(t)\right]\left|+\right\rangle\left|\alpha\right\rangle,\nonumber\\
U(t,t_{0})\left|-\right\rangle\left|\alpha\right\rangle&\approx&-g^{2}\left[aa^{\dag}G_{-}(t)+a^{\dag}aF_{-}(t)\right]\left|-\right\rangle\left|\alpha\right\rangle,
\label{A1}\nonumber
\end{eqnarray}
where $F_{\pm}(t)$ and $G_{\pm}(t)$ are given by Eq. (\ref{F}) and Eq. (\ref{G}), respectively, \begin{eqnarray}
F_{\pm}(t)&=&\int^{t}_{0} dt_{1}e^{\mp 2i \left(\int^{t_{1}}_{0}dt^{'}\Omega(t') \right)\mp i w t_{1}}\nonumber\\
&&\times\int^{t_{1}}_{0} dt_{2}e^{\pm 2i\left(\int^{t_{2}}_{0}dt^{''}\Omega(t'')\right)\nonumber
\pm i w t_{2}},\end{eqnarray}\begin{eqnarray}
G_{\pm}(t)&=&\int^{t}_{0} dt_{1}e^{\mp 2i \left(\int^{t_{1}}_{0}dt^{'}\Omega(t') \right)\pm i w t_{1}}\nonumber\\
&&\times\int^{t_{1}}_{0} dt_{2}e^{\pm 2i\left(\int^{t_{2}}_{0}dt^{''}\Omega(t'')\right)\mp i w t_{2}},\nonumber
\end{eqnarray} corresponding  to terms of second order of the operator $U(t,t_{0})$ from Eq. (\ref{10}).
\begin{figure}[h]
\begin{center}
\includegraphics[angle=0,width=2.8in]{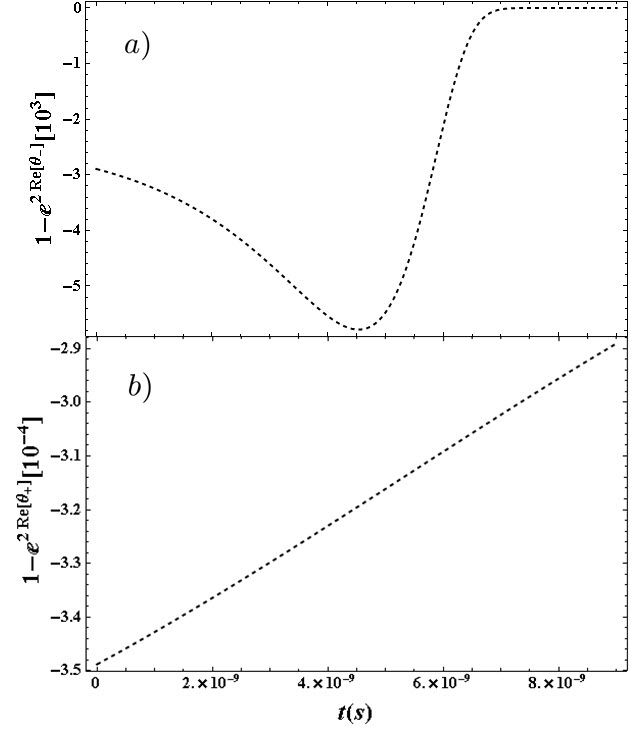}
\caption{The function $f_{\pm}(t)$ operating on the states $\left|\pm\right\rangle$.  } \label{re}
\end{center}
\end{figure}

The second member of the evolution equations above can be respectively rewritten as
\begin{eqnarray}
{-g^{2}\left\{G_{+}(t)+\left[F_{+}(t)+G_{+}(t)\right]a^{\dag}a\right\}\left|+\right\rangle\left|\alpha\right\rangle},\nonumber\\
{-g^{2}\left\{G_{-}(t)+\left[F_{-}(t)+G_{-}(t)\right]a^{\dag}a\right\}\left|-\right\rangle\left|\alpha\right\rangle},\nonumber
\label{}
\end{eqnarray}
or alternatively as
\begin{eqnarray}
{-g^{2}}{G_{+}(t)}\left\{1+\left[1+\frac{F_{+}(t)}{G_{+}(t)}\right]a^{\dag}a\right\}\left|+\right\rangle\left|\alpha\right\rangle,\\
{-g^{2}}{G_{-}(t)}\left\{1+\left[1+\frac{F_{-}(t)}{G_{-}(t)}\right]a^{\dag}a\right\}\left|-\right\rangle\left|\alpha\right\rangle.
\label{}
\end{eqnarray}
The terms $-g^{2}G_{\pm}(t)$ give only contributions to a global
phase of no implication and without any loss of generality are
being neglected.  For short time and small average number of
photons, $\left[1+\frac{F_{\pm}(t)}{G_{\pm}(t)}\right]a^{\dag}a$
is kept small enough so that the following approximation can be
employed
\begin{eqnarray}
1+\left[1+\frac{F_{\pm}(t)}{G_{\pm}(t)}\right]a^{\dag}a&\approx&\exp\left\{\left[1+\frac{F_{\pm}(t)}{G_{\pm}(t)}\right]a^{\dag}a\right\},\nonumber
\end{eqnarray} and thus
\begin{equation}
\left[1+ \theta_{\pm}(t) a^\dagger a \right]\left|\pm\right\rangle\left|\alpha\right\rangle\rightarrow e^{-\frac 1 2 |\alpha|^2f(t)}\left|\pm\right\rangle\left|\alpha
e^{\theta_{\pm}(t)}\right\rangle, \nonumber
\end{equation} as in Eq. (\ref{11})
where  $f_{\pm}(t)=[1-e^{2 \textrm{Re} (\theta_{\pm}(t))}]$,
$\theta_{\pm}(t)=1+\frac{F_{\pm}(t)}{G_{\pm}(t)}$. This
approximation has to be taken with caution.  As shown in Fig.
(\ref{re}a), $f_{-}(t)$ increases to large (negative) values,
achieving its maximal value around $4.5 \textrm{ns}$ meaning that
at those times the approximation in Eq.(\ref{11}) is not good
enough. However closer to the time of optimal phase accumulation,
$t_{op}=7.5 \textrm{ns}$, in Fig. (\ref{fig4}), $f_{-}(t)$
decreases rapidly to  ($\approx 10^{-3} -10^{-4}$), meaning that
the accumulated real part of the pulse is negligible well before
the end of the pulse.  On the other hand the function
$f_{+}(t)=1-e^{2\textrm{Re}\left\{\theta_{+}(t)\right\}}$, shown
in Fig. (\ref{re}b), is always very small, of order $ 10^{-3}
-10^{-4}$, during the time of the pulse. Thus showing the validity
of the approximation in Eq. (\ref{11}), is consistent for the time
of the pulse.

Now at the optimal time, $t_{op}$, since $f_{\pm}(t_{op})\approx 0$, Eq. (\ref{11}) can be effectively written as
\begin{equation}
\left[1+ \theta_{\pm}(t_{op}) a^\dagger a \right]\left|\pm\right\rangle\left|\alpha\right\rangle\approx \left|\pm\right\rangle\left|\alpha
e^{\theta_{\pm}(t_{op})}\right\rangle.
\end{equation}

\end{document}